\documentclass{nature}

\usepackage{amsmath}
\usepackage[pdftex]{graphicx}
\usepackage{dcolumn}
\usepackage{bm}     
\usepackage{dsfont}
\usepackage{color}
\usepackage{braket}
\usepackage{hyperref}

\newcounter{defcounter}
\usepackage{amsthm}

\title{Does electronic coherence enhance anticorrelated pigment vibrations under realistic conditions?}

\author{Hong-Guang Duan$^{1,2,3}$, Michael Thorwart$^{2,3}$ \& R. J. Dwayne Miller$^{1,3,4}$} 

\begin{document} 

\maketitle

\begin{affiliations}
\item Max Planck Institute for the Structure and Dynamics of Matter, Luruper
Chaussee 149, 22761, Hamburg, Germany
\item I.\ Institut f\"ur Theoretische Physik,  Universit\"at Hamburg,
Jungiusstra{\ss}e 9, 20355 Hamburg, Germany
\item The Hamburg Center for Ultrafast Imaging, Luruper Chaussee 149, 22761
Hamburg, Germany
\item The Departments of Chemistry and Physics, University of Toronto, 80 St.
George Street, Toronto Canada M5S 3H6\\
\centerline{\underline{\date{\bf \today}}}
\end{affiliations}

\begin{abstract}
The light-harvesting efficiency of a photoactive molecular complex is largely determined by the properties 
of its electronic quantum states. Those, in turn, are influenced by molecular vibrational states of the nuclear degrees of freedom. Here, we reexamine two recently formulated concepts that a coherent vibronic coupling between molecular states  would either extend the electronic coherence lifetime or enhance the amplitude of the anticorrelated vibrational mode at longer times. For this, we study a vibronically coupled dimer and calculate the nonlinear two-dimensional (2D) electronic spectra which directly reveal electronic coherence. The timescale of electronic coherence is initially extracted by measuring the anti-diagonal bandwidth of the central peak in the 2D spectrum at zero waiting time. Based on the residual analysis, we identify small-amplitude long-lived oscillations in the cross-peaks, which, however, are solely due to groundstate vibrational coherence, regardless of having resonant or off-resonant conditions. Our studies neither show an enhancement of the electronic quantum coherence nor an enhancement of the anticorrelated vibrational mode by the vibronic coupling  under ambient conditions.
\end{abstract} 

In the initial steps of photosynthesis, photoactive molecular complexes capture the sunlight energy and transfer it to the reaction center on an ultrafast time scale and with unity quantum efficiency \cite{Blankenship book}. The performance is determined by the molecular electronic properties, in concert with the molecular vibrations and coupling to the environment given by a solvent and the surrounding pigments and proteins. To investigate the energy transfer,  ultrafast 2D electronic spectroscopy  \cite{Ann Rev Phys Chem 54 425 (2003), CPL 386 184 2004, JCP 121 4221 2004} is able to resolve fs time scales. It is able to reveal the interactions between the energetically close-by lying molecular electronic states, for which the linear spectra are commonly highly congested and broadened by the strong static disorder \cite{Mukamel book}. Recent experimental studies of the Fenna-Matthews-Olson (FMO) complex reported long-lived oscillations of the cross-peaks both at low \cite{Nature 446 782 2007} and at room temperature \cite{PNAS 107 12766 2010} which have been assigned to enhanced electronic coherence. This has generated tremendous interest in this new  
 field of quantum biology \cite{Nat Phys 9 10 2013}, aiming to reveal a functional connection between photosynthetic energy transfer and long-lived quantum coherence. Moreover, also in photoactive marine cryptophyte algae \cite{Nature 463 644 2010}, the light-harvesting complex LHCII \cite{Nat Chem 4 389 2012} and in the Photosystem II reaction center \cite{Nat Chem 6 706 2014, Nat Phys 10 676 2014}, long-lived oscillations have been experimentally reported at low and room temperature. 

To model the reported \cite{Nature 446 782 2007} coherence, Ishizaki and Fleming have used a parametrized model of the FMO complex \cite{PNAS 106 17255 2009}, with a rather small reorganization energy of 35 cm$^{-1}$ to fit the electronic coherence timescale \cite{JCP 139 235102 2013}. This value was extracted \cite{Biophys J 91 2778 2006} from flourescence line narrowing measurements at low temperature \cite{Wendling} and does not include high-frequency intramolecular modes. However, even with the small reorganization energy, Shi {\em et al.} have calculated the complete 2D spectra and found a much shorter electronic coherence lifetime \cite{JCP 134 194508 2011}. They have pointed out that the interpretation of the long-lived coherence could just be due to the intentional magnification of the 2D spectral amplitudes by the deliberately used inverse hyperbolic sine scale. In addition, electronic quantum coherence has been questioned to play any crucial role for the energy transfer as the transport is domainted by largely incoherent exciton relaxation \cite{PRE 84 041926 2011,JCP 137 174111 2012}. A critical issue has been the use of an inadequate spectral distribution of the environmental fluctuations. The experimentally determined spectral density with a larger reorganization energy  \cite{Biophys J 91 2778 2006} has been used to calculate the dynamics by the quasiadiabatic propagator path integral \cite{PRE 84 041926 2011}. There, a local vibrational mode at 180 cm$^{-1}$ with a broadening of 29 cm$^{-1}$ has been included, with a total reorganization energy of 100  cm$^{-1}$. The numerically exact results also show a significantly shorter electronic coherence lifetime. Recent QM/MM-simulations \cite{Coker1,Coker2,Coker3,KK1,KK2} yield site-resolved spectral densities with reorganization energies of 150 to 200 cm$^{-1}$. 

Thus, theoretical studies showed that pure electronic quantum coherence can not survive under ambient conditions. Motivated by this disagreement, the coherent exciton dynamics in the FMO complex has been reexamined experimentally by 2D electronic spectroscopy \cite{PNAS 114 8493 2017}. A fit to an Ohmic spectral density with a broadened high-frequency mode yields a reorganization energy of 190 cm$^{-1}$. The observed  lifetime of the electronic coherence of $\sim$60 fs is too short to play any functional role in the energy transport, which occurs of the ps time scale. 

In addition to the electronic coherence, signatures of the vibrational coherence of the pigment-protein host can also be accessed on the same spectroscopic footing \cite{JPCA 118 10259 2014, CPL 545 40 2012, JPCC 117 18728 2013, JPClett 3 2828 2012}. Yet, electronic coherence can be distinguished from vibrational coherence \cite{JPCB 117 9380 2013,NJP 17 072002 (2015), Nat Chem 6 196 (2014)}. Long-lived pure electronic coherence is unexpected to exist in most light harvesting complexes. However, long-lived vibrational coherence is common and is not expected to strongly affect light harvesting in the first place. Two concepts are currently under debate: i) In 2013, Plenio {\em et al.} \cite{Nat Phys 9 113 2013} argued that long-lived vibrationally coherent modes can significantly enhance the electronic coherence lifetime when the vibrational and electronic degrees of freedom are resonantly coupled. The vibrational mode thereby is supposed to act as a ``phonon laser'' on the excitons, thereby producing ultralong electronic coherence. ii) Moreover,  Tiwari {\em et al.} \cite{PNAS 110 1203 (2013)} argued that nonadiabatic electronic-vibrational coherent mixing at short times resonantly enhance the amplitude of the particular delocalized anticorrelated vibrational mode on the ground electronic state. This second concept does not involve long-lived electronic coherence and is conceptually in agreement with the observation that a strong vibronic coupling produces large-amplitude coherent oscillations of the electronic component with a usual short lifetime and a long-lived vibrational coherence, but with a rather small amplitude \cite{NJP 17 072002 (2015), Nat Chem 6 196 (2014)}. 
This scenario was re-examined again recently \cite{KaisEngel2018} with an explicit coupling to an electronic and a vibrational bath. The result that an increased vibronic coupling survives weak electronic dephasing at short times and induces a resonantly enhanced long-lived vibrational coherence of the anticorrelated mode was presented as an explanation of the long-lived coherence signals observed in Ref.\  \cite{Nature 446 782 2007}. However, again, an unrealistically weak electronic damping with a reorganization energy of 35 cm$^{-1}$ has been used. 

Motivated by this discrepancy, we have reexamined the coherent dynamics in a model of a vibronically coupled excitonic dimer with anticorrelated pigment vibrations. We calculate the 2D electronic spectra of the model dimer as being part of the FMO complex. We use the environmental parameters obtained from the recent FMO experiment \cite{PNAS 114 8493 2017}. We examine the electronic coherence lifetime by the antidiagonal bandwith of the diagonal peak at zero waiting time. The oscillations in the residuals obtained from the global fitting analysis  confirm that the long-lived coherence is purely vibrational in nature, irrespective of resonant or non-resonant conditions. To distinguish the coherent dynamics of the electronic excited state, we have calculated the dynamics of the electronic wave-packet on the vibrational potential energy surfaces (PESs), accompanied by the projection onto the reaction coordinates. Also here, the vibrational coherence is clearly identified by the oscillations close to the potential minimum. The projection  shows that the long-lived oscillations are solely of vibrational origin, which confirms the 2D spectroscopic calculations. Moreover, we show that under realistically strong electronic damping, coherent vibronic coupling at short times does not enhance the amplitude of the anticorrelated vibrational mode, while we recover the mechanism of vibronic enhancement only for unrealistically weak electronic damping. 

\section*{Theory}
The model is described by a total Hamiltonian consisting of the system, bath and the system-bath interaction terms, $H=H_{S}+H_{SB}$. The system is a dimer consisting of monomer A and B with site energies $E_{A/B}$, both having the same electronic ground state $\ket{g}$ and the respective electronic single excited states $\ket{A}$ and $\ket{B}$. The double excited state is denoted as $\ket{AB}$. Each electronic excited state couples to a vibrational mode (each to its own mode). The two couplings are such that an anticorrelated out-of-phase oscillation of the two electronic states occurs. In the exciton site basis, we have
\begin{eqnarray}
\label{eq:sysham}
H_{S}& =&\ket{g}h_{g}\bra{g} + \ket{A}h_{A}\bra{A} + \ket{B}h_{B}\bra{B} \nonumber \\
 &&+(\ket{A}V\bra{B}+h.c.) + \ket{AB}(h_{A}+h_{B})\bra{AB} \, .
\end{eqnarray}
Here, $h_{g}=\frac{1}{2}\Omega(P^2_{A/B}+Q^2_{A/B})$, $h_{A}=E_{A}+h_{g}-\kappa Q_{A}$ and $h_{B}=E_{B}+h_{g}+\kappa Q_{B}$, respectively. $P_{A/B}$ and $Q_{A/B}$ are the momenta and the coordinates of the two vibrational modes coupled to  monomer A and B. We express the vibronic coupling strength between ground and excited state as  $\kappa=\frac{\Omega \Delta }{\sqrt 2}$, where $\Delta$ is the dimensionless shift of the excited state relative to its ground state. $\Omega$ is the vibrational frequency (both modes are taken with equal characteristics). $V$ denotes the electronic coupling between two electronic excited states $\ket{A}$ and $\ket{B}$.

For the discussion of the anticorrelated vibrations, it is useful to define new coordinates and momenta according to $Q_{\pm}=\frac{1}{\sqrt{2}}(Q_{A}\pm Q_{B})$, and $P_{\pm}=\frac{1}{\sqrt{2}}(P_{A}\pm P_{B})$ \cite{JCP 136 034507 (2012)}. Then, the system Hamiltonian can be written as 
%
%
%
\begin{eqnarray}
\label{eq:totalham}
  H_{S} &= & \sum_{n=0,1,2}H_{S,n} = \sum_{n=0,1,2}(H^{+}_{S,n}+H^{-}_{S,n}), \nonumber \\
  H^{-}_{S,0} &= &\mathbf{1}_{0}h^{-}, \nonumber \\
  H^{-}_{S,1} &= &\mathbf{1}_{1}h^{-}+(E_{B}-\Omega Q_{-} \Delta/\sqrt{2}) \ket{B}\bra{B} \nonumber\\
                &&+(E_{A}+\Omega Q_{-}\Delta/\sqrt{2})\ket{A}\bra{A}\nonumber\\
                &&+V(\ket{A}\bra{B}+\ket{B}\bra{A}), \nonumber\\
  H^{-}_{S,2}& =& \mathbf{1}_{2}(h^{-}+E_{A}+E_{B})\, ,
\end{eqnarray}
with $[H^{+}_{S,n},H^{-}_{S,n}] = 0$ and $H^{+}_{S,n} = \mathbf{1}_{n}(h^{+}-n \Omega Q_{+}\Delta/\sqrt{2})$. 
The two rotated vibrational modes are given by $h^{\pm}=\frac{1}{2}\Omega(P^2_{\pm}+Q^2_{\pm})$ and the projection operators are  $\mathbf{1}=\sum_{n=0,1,2}\mathbf{1}_{n}$, with $\mathbf{1}_{0}=\ket{g}\bra{g}$, $\mathbf{1}_{1}=\ket{A}\bra{A}+\ket{B}\bra{B}$, and $\mathbf{1}_{2}=\ket{AB}\bra{AB}$. 

We choose the same parameters as in Ref.\ \cite{PNAS 110 1203 (2013)} for the system Hamiltonian. This dimer mimics one exciton pair of the FMO complex. The bath part will be discussed below (in particular, we do not choose the same parameters of too weak damping, but use our own parameters of Ref.\ \cite{PNAS 114 8493 2017}). The electronic energy gap is set to $E_{A}-E_{B}=150$ cm$^{-1}$ and the electronic coupling is $V=66$ cm$^{-1}$. Moreover, the dimensionless vibrational shift is set to $\Delta=0.2236$. The pure electronic energy gap without coupling can then be calculated to be $\Delta E=200$ cm$^{-1}$. As in Ref.\  \cite{PNAS 110 1203 (2013)}, 
we model inhomogeneous broadening by a static Gaussian disorder of width $\delta E=26$ cm$^{-1}$. For the vibrational mode, we choose the frequency $\Omega=200$ cm$^{-1}$ which corresponds to the resonant case when the vibronic coupling vanishes (the slight shift of this resonance due to the vibronic coupling  does not alter the overall result since we find essentially the same conclusion also for the off-resonant case, see below). 


The environment and coupling part $H_{SB}=H_{SB}^{\rm vib}+H_{SB}^{\rm el}$ consist of two parts, the vibrational baths which damp the vibrational motions, and the electronic bath which generates electronic dephasing and damping. In general, we assume Gaussian fluctuations described in terms of the standard model of dissipative quantum systems. The electronic environment is generated by fluctuating charges in the protein and the solvent and consists of two  harmonic oscillator baths each of which couples to the electronic excited states of monomer A and B, respectively. Thus, we have the Hamiltonian $H_{SB}^{\rm el}=\sum_{\alpha=A,B}H_{SB,\alpha}^{\rm el}$ with 
\begin{equation}
\label{eq:bath_ham}
H_{SB,\alpha}^{\rm el}  = \frac{1}{2}\sum^{N}_{i=1}\left[\frac{p^{2}_{i,\alpha}}{m_{i,\alpha}}  + m_{i,\alpha}\omega^{2}_{i,\alpha}\left(x_{i,\alpha}-
\frac{c_{i,\alpha}\ket{\alpha}\bra{\alpha}}{m_{i,\alpha}\omega^2_{i,\alpha}}
 \right)^2
 \right] . 
\end{equation}
As usual, $p_{i,\alpha}$ and $x_{i,\alpha}$ are the momenta and the coordinates of the $i$th bath mode coupling to the electronic state $\alpha=A,B$. For the electronic part, we choose an Ohmic spectral density with the parameters obtained from fitting the linear spectra of the FMO complex to experimental data, see Ref.\ \cite{PNAS 114 8493 2017}. Notice that these values  correspond to much stronger damping than those in Ref.\ \cite{PNAS 110 1203 (2013)}. Thus, each bath  is assumed to have its own, but equal spectral density $J^{\rm el}(\omega)=\gamma^{\rm el}\omega\exp(-\omega/\omega_{c})$, 
with $\gamma^{\rm el}=0.7$, $\omega_{c}=350$ cm$^{-1}$. 

The vibrational environment roots in fluctuating nuclear degrees of freedom of the protein and couples to the vibrational displacements $Q_A$ or $Q_B$ of the mode coupled to the electronic state $A$ or $B$. Hence, $H_{SB}^{\rm vib}=\sum_{\alpha=A,B}H_{SB,\alpha}^{\rm vib}$ with 
\begin{equation}
\label{eq:bath_ham}
H_{SB,\alpha}^{\rm vib}  = \frac{1}{2}\sum^{N}_{i=1}\left[\frac{q^{2}_{i,\alpha}}{\mu_{i,\alpha}} + \mu_{i,\alpha}\nu^{2}_{i,\alpha} \left(y_{i,\alpha}-
\frac{d_{i,\alpha}Q_\alpha}{\mu_{i,\alpha}\nu^2_{i,\alpha}}
  \right)^2
 \right] . 
\end{equation}
$q_{i,\alpha}$ and $y_{i,\alpha}$ are the momenta and the coordinate of the $i$th vibrational bath mode of the state $\alpha=A,B$. We assume that the vibrational bath has the same spectral density as the electronic bath, i.e.,  $J^{\rm vib}(\omega) = \gamma^{\rm vib}\omega\exp(-\omega/\omega_{c})$ but with weaker damping, $\gamma^{\rm vib}=0.02$ and $\omega_{c}=350$ cm$^{-1}$. 

To disentangle  electronic and vibrational coherence, we perform a projection of the electronic wave packet on the reaction coordinate, which allows us to distinguish the vibrational coherence from the vibronic dynamics. We assume the initial wave packet to be in the lowest vibrational state $\ket{0}$ of the electronic excited state $\ket{A}$ in the site basis, such that the initial density matrix can be written as $\rho(0)=\ket{A,0}\bra{A,0}$. In order to obtain dynamical information, we determine the probability of the wave packet along the reaction coordinate $Q_{-}$ by the time-dependent projection 
\begin{equation}
\label{eq:bath_ham}
P^{ad}_{k}(Q_{-},t) = \bra{Q_{-}}\bra{\tilde{k}}\rho(t)\ket{\tilde{k}}\ket{Q_{-}} \, , 
\end{equation}
where $P^{ad}_{k}$ is the probability density of the reaction coordinate and $\tilde{k}$ indicates the electronic state of A or B in the exciton basis  (for details of the projection, see Refs.\ \cite{JCP 93 345 (1990), JCP 147 074101 (2017)}).

\section*{Results and Discussion}
We assume \cite{PNAS 110 1203 (2013)} that two perpendicular transitions from the common ground state to the two excited states of monomer A and B are possible. Hence, the transition dipole moments are fixed to $\vec{\mu}_{A}=\mu_{A}\mathbf{e}_{x}$ and $\vec{\mu}_{B}=\mu_{B}\mathbf{e}_{y}$ with $\mu_{A}=\mu_{B}=1$. Here, $\mathbf{e}_{j}$ is the unit vectors in the direction $j$. Temperature is set to $300$ K, if not stated otherwise. We use the time non-local quantum master equation \cite{JCP 111 3365 (1999), JCP 121 2505 (2004), JPCA 112 4254 (2008)} with the equation-of-motion phase-matching-approach \cite{JCP 123 164112 (2005)}. Details are given in the Supplementary Information Appendix of Ref.\ \cite{PNAS 114 8493 2017}. 

\begin{figure}[h!]
\begin{center}
\includegraphics[width=17.0cm]{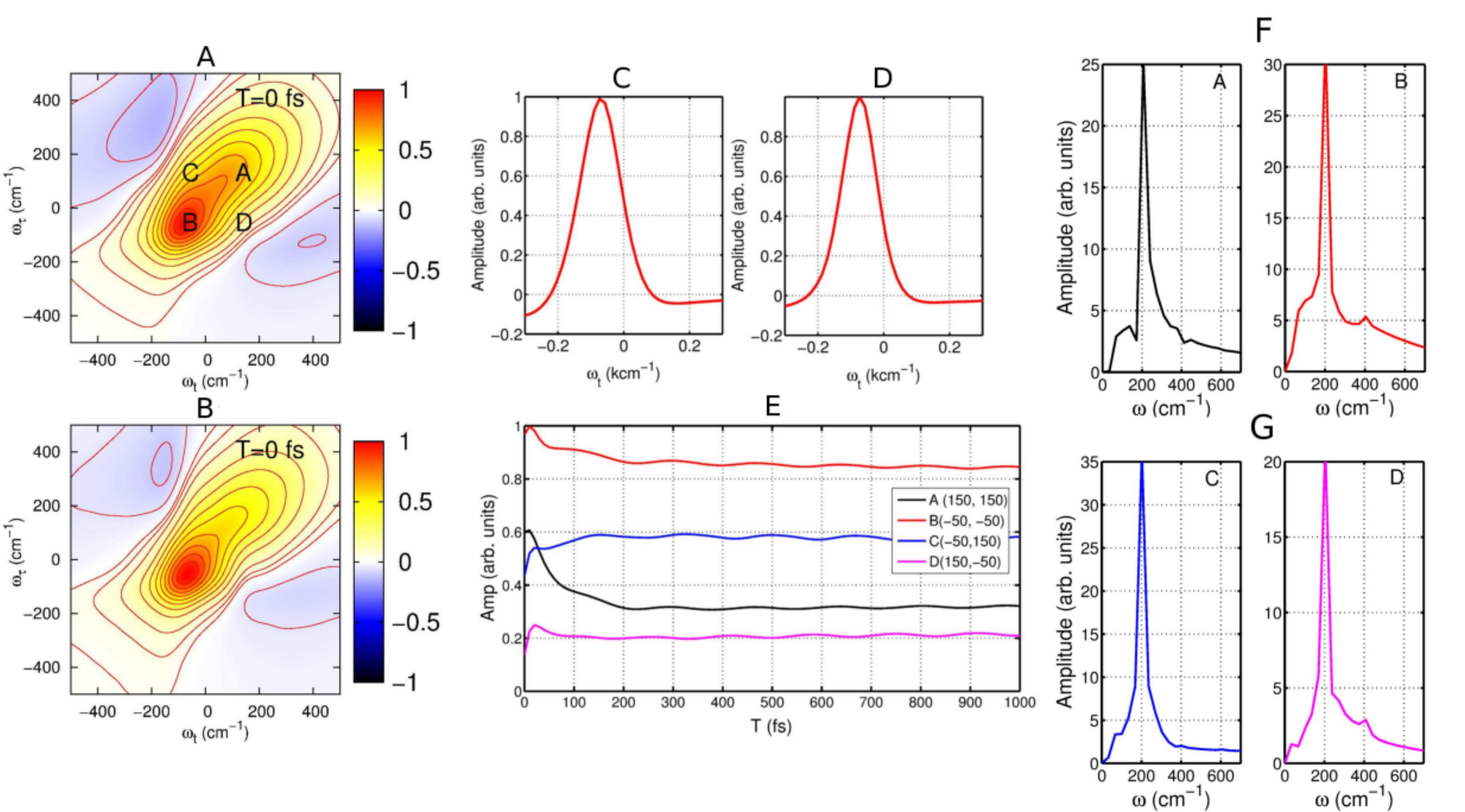}
\caption{\label{fig:Fig1} (A) Real part of the 2D electronic spectrum of the vibronic dimer at room temperature (300 K). The diagonal peak A is located at ($\omega_{t}=150$ cm$^{-1}$, $\omega_{\tau}=150$ cm$^{-1}$) and B at ($\omega_{t}=-50$ cm$^{-1}$, $\omega_{\tau}=-50$ cm$^{-1}$). The off-diagonal peak C sits at ($\omega_{t}=-50$ cm$^{-1}$, $\omega_{\tau}=150$ cm$^{-1}$) and D at ($\omega_{t}=150$ cm$^{-1}$, $\omega_{\tau}=-50$ cm$^{-1}$). The time dependent trace of the selected peaks are shown in (E). For comparison, the 2D spectrum of the dimer without the effective vibrational mode is shown in (B). To obtain the timescale of the electronic dephasing, the anti-diagonal profile of the peaks B in (A) and (B) are shown in (C) and (D). The resulting timescale of the electronic dephasing is 70 fs and 80 fs, respectively. (F), (G) Power spectra of the peaks A, B, C, and D for the case with vibrational coupling. The broad background spectral band is associated to fast electronic dephasing. In addition to the strong vibrational peak at 200 cm$^{-1}$, one additional peak at 400 cm$^{-1}$ is well resolved. }
\end{center}
\end{figure}

We first consider the vibronic dimer with resonant vibrational coupling for which we obtain the 2D electronic spectrum shown in Fig.\ \ref{fig:Fig1}(A) (real part). At waiting time $T=0$ fs, the inhomogeneous broadening can be clearly identified because the spectrum is stretched along the diagonal. It disappears within $50$ fs (see the time dependent 2D electronic spectra in the SI). The corresponding 2D spectrum of the purely electronic dimer without vibrational modes is  shown in Fig.\ \ref{fig:Fig1}(B). From the profiles taken along the antidiagonal band and shown in Figs.\ \ref{fig:Fig1} (C) and (D), we observe that the anti-diagonal broadening of the 2D spectra in both cases is similar, which proves that both dimers undergo a similar dephasing dynamics with similar time scales of the electronic dephasing. They are extracted to be 70 fs and 80 fs, respectively. To resolve the time-dependent energy transfer, we analyze the series of 2D spectra for increasing waiting times, by the global fitting approach, see Supplementary Information Appendix for details. This yields the shortest lifetime of the decay associated spectra which is induced by the peak broadening and by electronic dephasing. Both lifetimes coincide in both cases which shows that a vibronic coupling does not alter the short-time electronic dephasing properties.

For a quantitative analysis of the dissipative dynamics in the presence of a vibrational coupling, we plot in Fig.\ \ref{fig:Fig1} (E) the time evolution of the magnitude of the peaks selected in Fig.\ \ref{fig:Fig1} (A). We observe that the dynamics can be clearly separated into two sectors: (i) fast electronic dephasing, which initially occurs on the time scale of $\sim$70 fs, as already resolved by the analysis of the anti-diagonal bandwidth and the global fitting approach. Moreover, (ii) long-lived oscillations with small amplitudes are resolved at longer times. In order to identify the origin of these long-lived oscillations, we perform a Fourier transform of the residual, which is obtained by subtracting the kinetics resolved by the global fitting. The result is shown in Figs.\ \ref{fig:Fig1} (F) and (G). The process of fast electronic dephasing is associated to the broad background spectral band with a maximum at 200 cm$^{-1}$. In addition, this broad band is overlapped by one sharp peak at the same frequency of 200 cm$^{-1}$. One additional narrow peak is resolved at $400$ cm$^{-1}$. It originates from the vibrational coherence between the vibrational ground $\ket{0}$ and the second vibrational level $\ket{2}$ of the electronic ground state, i.e., is of pure vibrational origin. This is further illustrated in Fig.\ S3 of the Supplementary Information Appendix, where the 
 stick spectrum also indicates the clearly separated electronic and vibrational parts of which the eigenstates are composed. Hence, as a matter of fact, we can conclude that these narrow peaks at 200 cm$^{-1}$ and 400 cm$^{-1}$ only stem from  vibrational coherence of each monomer. 

\textbf{Off-resonance case} Up to here, we have studied the vibronic dimer for the resonant case $\Delta E = \Omega$. Next, we investigate the off-resonant case as well and choose $\Omega=500$ cm$^{-1}$. The results are shown in Fig.\ \ref{fig:Fig3} (A), the global fitting analysis is shown in the Supplementary Information Appendix. The fast electronic dephasing  with the time scale of $69$ fs is still present. It agrees with the value of the antidiagonal bandwidth (see the SI). Importantly enough, it coincides with the dephasing time scale of the resonant case. Hence, the fact that the electronic and vibrational dynamics are off-resonant  does not affect the conclusion reached for the resonant case. In addition, we show in Fig.\ \ref{fig:Fig3} (B) the dynamics of the selected peaks for growing waiting times. It shows the same kinetics as in the resonant case: One fast electronic dephasing component is combined with a long-lived vibrational coherent component with a small amplitude. Again, we perform the Fourier transform of the residuals and plot the  spectra of each peak in Fig.\ \ref{fig:Fig3} (C). We again find one broad band with a maximum at $200$ cm$^{-1}$, which manifests the fast electronic dephasing and coincides with the lifetime of $\sim70$ fs resolved by the global fitting approach. One clearly separated narrow peak is located at $500$ cm$^{-1}$ with a large magnitude which is associated to the long-lived vibrational coherence. A clear evidence for the purely vibrational (and not vibronic) origin of the peak is that one additional peak can be resolved at $\sim1000$ cm$^{-1}$. It is the clear signature of the vibrational coherence between the vibrational ground $\ket{0}$ and the second vibrationally-excited level $\ket{2}$ on the electronic ground-state surface. 

\textbf{Low-temperature case} Next, we consider the case of low temperature of $80$ K. We follow the same steps as above and find for vanishing vibronic coupling $\kappa=0$ an electronic dephasing time scale of $161$ fs, see Fig.\ S4 in the SI. In the presence of a resonant vibrational mode with $\Omega=200$ cm$^{-1}$ and with $\kappa \ne 0$ set to the same value as above, we obtain the dephasing time of $121$ fs, see Fig.\ S3 (B) in the SI. It is slightly smaller than the one of the purely electronic case, but still comparable. The dynamics of the selected peaks in the 2D spectra  again shows long-lived oscillations, see Fig.\ S3 (C) of the SI. The Fourier transform, shown in Fig.\ S3 (D), shows again one sharp peak at $200$ cm$^{-1}$, and one additional peak at $400$ cm$^{-1}$ with a quite weak magnitude. They manifest again the vibrational origin of the coherence. Therefore, we can conclude that, also for low temperature, the long-lived oscillation is just of vibrational origin. No  different mechanism between low and room temperature occurs. 

\textbf{Vibrational dynamics of the monomer} 
In addition to the dimer, we also investigate the monomer where only vibrational coherence is present. In Fig.\ S5 (B), we show the time trace of the selected cross peak together with the Fourier spectrum in (C). The spectra are dominated by one peak at the vibrational frequency. An  additional peak appears at the position of twice the vibrational frequency. Thus, the same scenario occurs for the monomer as well. We clearly demonstrate that the long-lived oscillations in a vibronically-coupled dimer are just due to the overlap of the short-lived electronic coherence and the long-lived vibrational coherence. 

\textbf{Wave packet tracking} 
A further confirmation of this picture is obtained from monitoring the dynamics of the electronic excited states. For this, we project the time-evolved density matrix onto the anticorrelated vibrational coordinate $Q_{-}$. We use the same parameters as before and calculate the PESs of the electronic excited states $\ket{\tilde{A}}$ and $\ket{\tilde{B}}$ in the adiabatic basis. The result for the off-resonant case with $\Omega=500$ cm$^{-1}$ is shown in Figs.\ \ref{fig:Fig4} (A) to (D). The initial wave packet is prepared in the excited state $\ket{A}$. For growing time, the transfer of population from $\ket{\tilde{A}}$ to $\ket{\tilde{B}}$ can be clearly identified by the decrease of the magnitude in Fig.\ \ref{fig:Fig4} (A) and the corresponding growth in Fig.\ \ref{fig:Fig4} (B). By this, the vibrational coherence of the excited states is clearly visible from the oscillations around the potential minimum, see Fig.\ \ref{fig:Fig4} (A). The oscillations have a period of $\sim66$ fs, which exactly coincides with the assigned vibrational frequency of 500 cm$^{-1}$. Moreover, the population dynamics of the states $\ket{\tilde{A}}$ and $\ket{\tilde{B}}$ is shown in Fig.\ \ref{fig:Fig4} (C) by summing the wave packet population along the reaction coordinate $Q_{-}$. Spectral information can be again obtained from the Fourier transform. In Fig.\ \ref{fig:Fig4} (D), the vibrational coherence is identified by the narrow peaks at 500 cm$^{-1}$ and 1000 cm$^{-1}$, which coincide with the results from the 2D spectroscopic calculations shown in Fig.\ \ref{fig:Fig3}. In addition, a broadband background with a maximum at 200 cm$^{-1}$ and with small magnitude is visible, which again provides evidence of the electronic coherence being short-lived. 

The resonant case, with $\Omega= 200$ cm$^{-1}$, is addressed in Fig.\ \ref{fig:Fig4} (E) to (F). Compared to the off-resonant case, no significant difference occurs. The initial wave packet in the excited state $\ket{\tilde{A}}$ is transferred to $\ket{\tilde{B}}$ over time. The only difference is the vibrational oscillation period of $\sim$165 fs. The integrated time-dependent populations are shown in Fig.\ \ref{fig:Fig4} (G) and the associated spectral information in Fig.\ \ref{fig:Fig4} (H). One narrow peak at 200 cm$^{-1}$ and one additional peak at 400 cm$^{-1}$  with quite small magnitude occur. Also here, the result agrees with the observation of the 2D spectroscopic calculations in Fig.\ \ref{fig:Fig1}.

\textbf{Vibronic dimer under weak electronic dephasing}
Up to here, we have studied realistic parameters of the electronic dephasing and the vibrational damping constants. The possibility remains that for weaker electronic dephasing,  the role of a coherent vibronic coupling could be more pronounced.   That this is not the case follows from the dynamics of a vibronic dimer in the off-resonant case with $\Omega=500$ cm$^{-1}$ under (unrealistically) weak electronic dephasing. For this, we set $\gamma^{\rm el}=\gamma^{\rm vib}=0.02$, and $\omega_{c}=50$ cm$^{-1}$. The wave packet dynamics projected to the PESs of $\ket{\tilde A}$ and $\ket{\tilde B}$ is shown in Fig.\ \ref{fig:Fig5} (A) and (B), respectively. The purely vibrational coherence can be seen from the wave packet oscillations around $Q_{-}=-1.5$ with a period of $\sim67$ fs, which coincides with the vibrational period. The electronic coherence is visible in Fig.\ \ref{fig:Fig5}  as a large-amplitude population exchange between the two electronic states. The electronic oscillation period of $\sim$167 fs corresponds to the electronic energy gap  $\Delta E=200$ cm$^{-1}$ in the adiabatic basis. Thus, the large-amplitude exchange is caused by the superposition of the wave packet components on the two PESs. To reveal the oscillation components and their lifetimes, we sum the wave packet components along the reaction coordinate and plot it in Fig.\ \ref{fig:Fig5} (C). The Fourier spectrum is shown in Fig.\ \ref{fig:Fig5} (D). Two large peaks at $200$ cm$^{-1}$ and $500$ cm$^{-1}$ correspond to the oscillations due to electronic and vibrational coherence, respectively. Two small vibronic peaks at $500-200=300$ cm$^{-1}$ and $500+200=700$ cm$^{-1}$ are due to the vibronic mixing. Most importantly, although the frequencies indeed mix and additional peaks are generated, the line widths of the peaks at $200$ cm$^{-1}$, $300$ cm$^{-1}$, $500$ cm$^{-1}$, and $500$ cm$^{-1}$ are $35$ cm$^{-1}$, $40$ cm$^{-1}$, $20$ cm$^{-1}$, and $45$ cm$^{-1}$, and are thus all comparable. This proves that the lifetime of the electronic coherence is not affected by vibronic coupling to a vibrational mode. 

\textbf{Impact of coherent vibronic coupling on anticorrelated vibrations}
Finally, we address the possibility that a strong coherent vibronic coupling could enhance the amplitude of the anticorrelated component of the vibrational dynamics \cite{PNAS 110 1203 (2013),KaisEngel2018}. The latter is given by the magnitude of the vibrational peak in the Fourier spectrum of the wave-packet dynamics, as, e.g., shown in Figs. \ref{fig:Fig4} D and H and Fig.\ \ref{fig:Fig5}D. In Fig.\ \ref{fig:Fig6}, we show the amplitude of the anticorrelated vibration for increasing vibronic coupling $\Delta$ for weak ($\gamma^{\rm el}=0.02$) and strong ($\gamma^{\rm el}=0.7$) electronic dephasing.  For increasing $\Delta$, the mixing of the anticorrelated vibration with the electronic parts becomes stronger. Indeed, for weak electronic dephasing, we find an increase of the anticorrelated vibrational amplitude which confirms the picture of Refs.\ \cite{PNAS 110 1203 (2013),KaisEngel2018}. However, for the more realistic case of stronger electronic dephasing, the amplitude of the anticorrelated vibration depends only  weakly on the vibronic coupling, since the coherent electron-vibrational mixing is dephased very rapidly and does not influence the vibration at later times.

%

\section*{Conclusions} 
In conclusion, we have shown that, under ambient physical conditions, it is irrelevant for the lifetime of electronic quantum coherence in the excitation energy transfer whether the two exciton states couple to two anticorrelated or correlated vibrational modes. This holds irrespective of whether the electronic and the vibrational transitions are resonant or off-resonant and follows from an analysis of a model dimer in which each excited state is coupled to its own vibrational mode in an anticorrelated manner. Two independent baths for electronic dephasing as well as vibrational damping are included. By this, we answer a key question in the literature \cite{Nat Phys 9 113 2013} whether a coupling to a long-lived vibrational mode can lead to a substantial increase of the electronic coherence time. The conclusions are drawn from the calculated dynamics, the 2D electronic spectra and the subsequent 2D global fitting approach. The exciton dynamics is characterized by the combination of a fast electronic dephasing  and a long-lived vibrational coherent component, which has very small oscillation amplitudes. 
The long-lived oscillations  are solely due to the coherence between different vibrational levels, irrespective of 
a resonant or an off-resonant vibronic anticorrelated coupling. Even under (unrealistically) weak electronic dephasing, the electronic coherence lifetime is not enhanced by the vibronic components. The same conclusion has been drawn from the study of indocarbocyanine dye molecules \cite{NJP 17 072002 (2015)}.  In addition, we find that a strong mixing of electronic and anticorrelated vibrational components of the wavefunction due to strong vibronic coupling does not enhance the vibrational amplitude at long times under ambient conditions. This effect only can occur under unrealistically weak electronic dephasing \cite{PNAS 110 1203 (2013),KaisEngel2018}, but does not play a role in realistic physical systems, the reason being that the required coherent vibronic mixing is rapidly destroyed by fast electronic dephasing. 
%



%
\begin{addendum}
 \item We acknowledge financial support by the Max Planck Society and the
Hamburg Centre for Ultrafast Imaging (CUI) within the German Excellence
Initiative supported by the Deutsche Forschungsgemeinschaft. H-G.D.
acknowledges generous financial support by the Joachim-Hertz-Stiftung Hamburg.

\item[Significance Statement] We have studied the impact of molecular vibrations on the electronic coherence during the energy transfer in a model dimer. The full dynamics are revealed in the calculated 2D electronic spectra. We show that the long-lived coherence present in the off-diagonal spectral signals is solely due to vibrational coherence in the monomer. Our calculations illustrate that neither the electronic coherence between two monomers can be enhanced by vibrations of individual pigments, irrespective of resonant or off-resonant conditions, nor can a coherent vibronic coupling enhance the amplitude of the anticorrelated vibrational mode under realistic conditions.

\item[Supporting Information] The Supplementary Information includes the global fitting approach and results, low-temperature calculations and vibrational dynamics of monomer.

\item[Competing Interests] The authors declare that they have no competing financial interests.

\item[Correspondence] michael.thorwart@physik.uni-hamburg.de and dwayne.miller@mpsd.mpg.de 
\end{addendum}
%

\newpage
\begin{figure}[h!]
\begin{center}
\includegraphics[width=17.0cm]{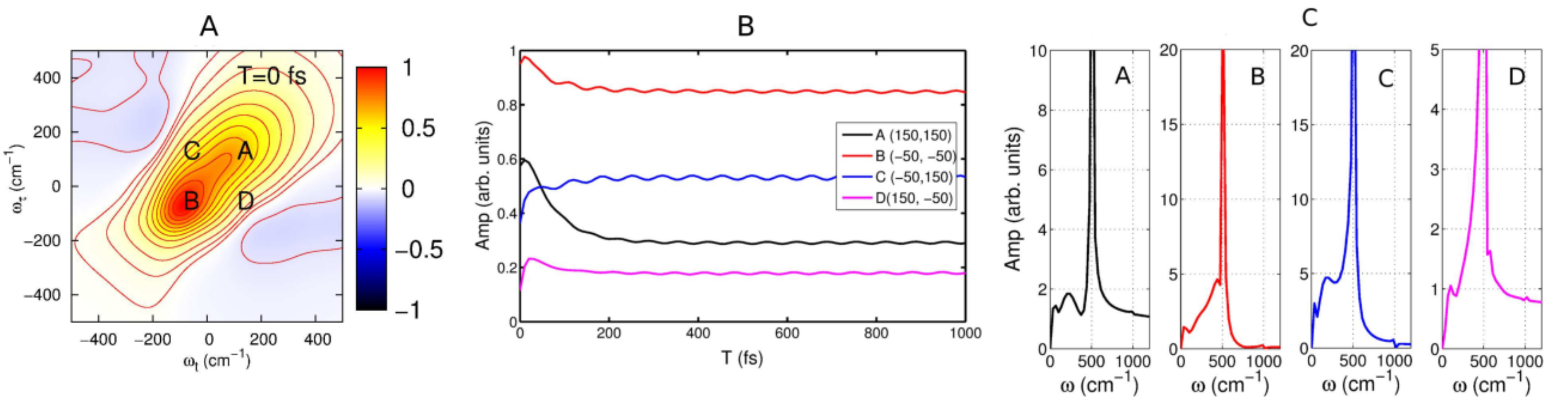}
\caption{\label{fig:Fig3} (A) Real part of the 2D electronic spectrum of the vibronic dimer under off-resonant conditions with $\Omega=500$ cm$^{-1}$ at $T=0$ fs. The kinetics of the selected peaks at A, B, C, and D are shown in (B), the associated power spectra are shown in (C). To resolve the lifetime of the electronic coherence, we fit the broad peak at 200 cm$^{-1}$ with the Lorentzian lineshape and obtain the  electronic coherence time of $\sim$70 fs, which agrees with the resonant case.}
\end{center}
\end{figure}

\newpage
\begin{figure}
\begin{center}
\includegraphics[width=17.0cm]{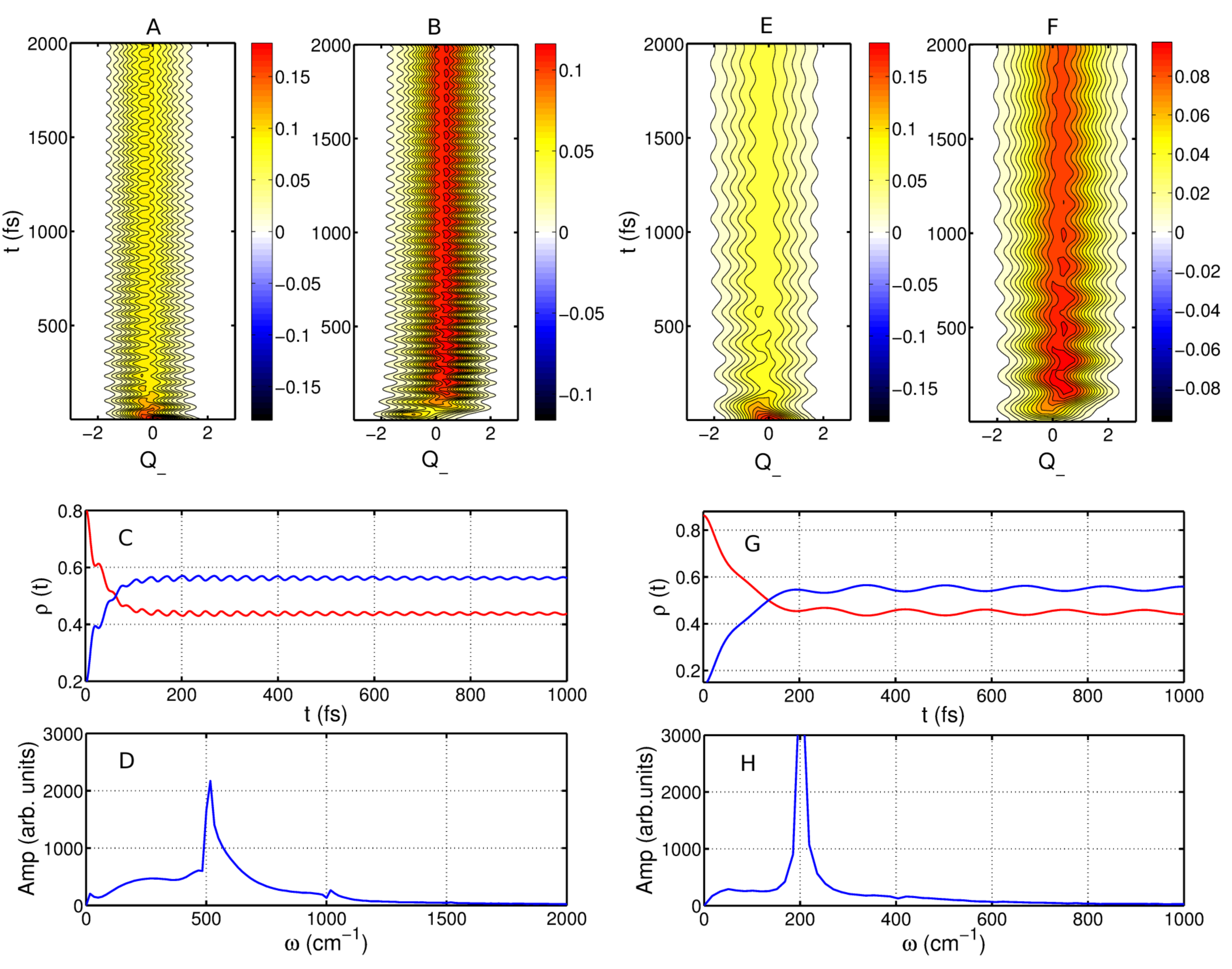}
\caption{\label{fig:Fig4} Time evolution of the wave-packet in the excited states $\ket{\tilde{A}}$ (A) and $\ket{\tilde{B}}$ (B). The integrated populations  of the electronic states $\ket{\tilde{A}}$ and $\ket{\tilde{B}}$ obtained by summing along the reaction coordinate $Q_{-}$ are shown in (C), together with the Fourier transform of the residuals in (D).  The long-lived vibrational coherence is identified by the narrow peaks at 500 cm$^{-1}$ and 1000 cm$^{-1}$. In addition, one broadband peak at 200 cm$^{-1}$ represents short-lived electronic coherence. The corresponding results of the resonant case are shown in (E) to (H), respectively.  The long-lived vibrational coherence can be identified by the narrow peak and 200 cm$^{-1}$ and the small peak at 400 cm$^{-1}$.}
\end{center}
\end{figure}

\newpage
\begin{figure}
\begin{center}
\includegraphics[width=17.0cm]{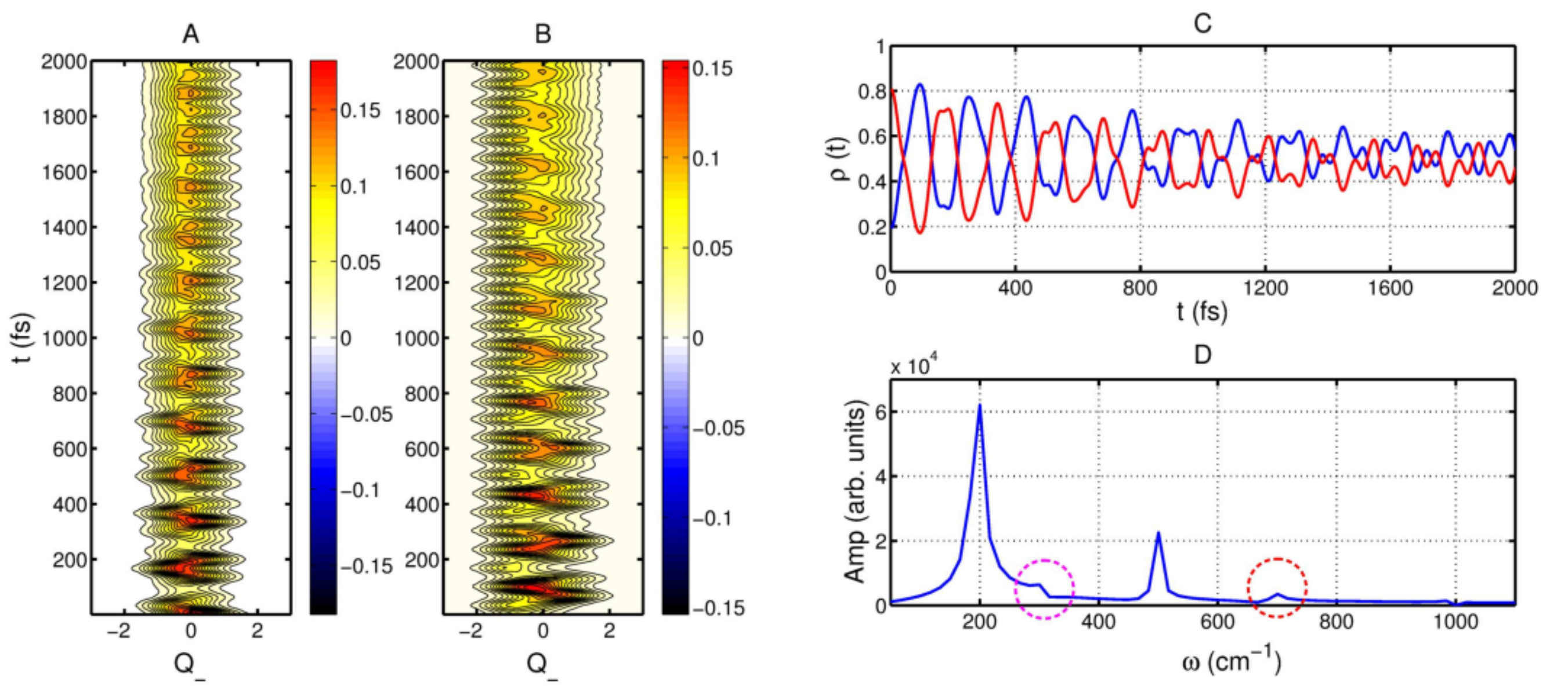}
\caption{\label{fig:Fig5} Time evolution of the wave-packet on the excited state PESs for $\ket{\tilde{A}}$ is shown in (A) and for $\ket{\tilde{B}}$ in panel (B) for the vibronic dimer under off-resonant conditions for very weak electronic dephasing $\gamma^{\rm el}=\gamma^{\rm vib}=0.02$, and $\omega_{c}=50$ cm$^{-1}$. The integrated populations  of the electronic states $\ket{\tilde{A}}$ and $\ket{\tilde{B}}$ obtained by summing along the reaction coordinate $Q_{-}$ are shown in (C), together with the Fourier transform of the residuals shown in (D).  The vibronic coherence can be identified by the two peaks $300$ cm$^{-1}$ and  $700$ cm$^{-1}$, which are marked by circles. }
\end{center}
\end{figure}

\newpage
\begin{figure}
\begin{center}
\includegraphics[width=10.0cm]{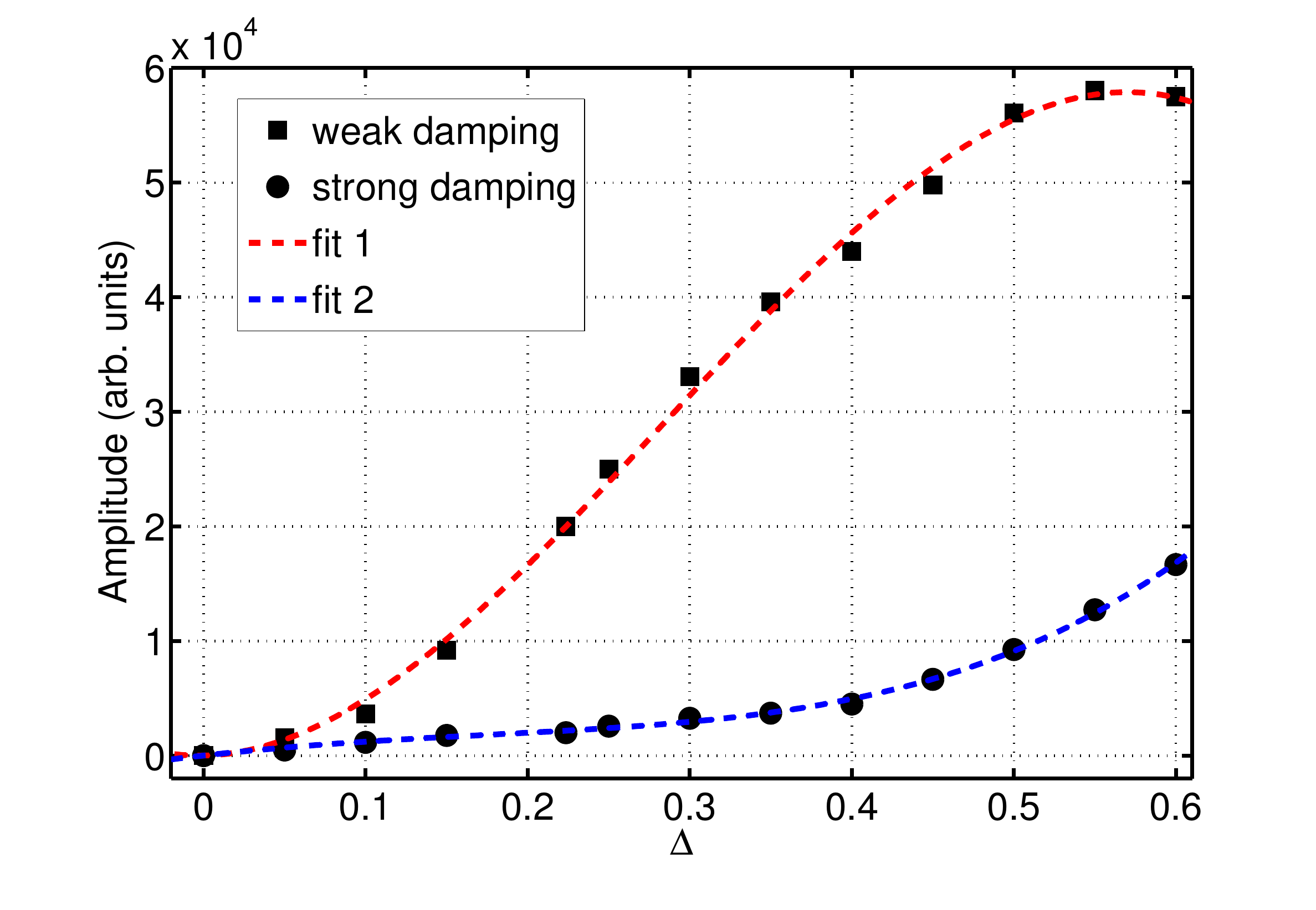}
\caption{\label{fig:Fig6} Oscillation amplitude of the anticorrelated vibration vs.\ the vibronic coupling strength  $\Delta$ for weak ($\gamma^{\rm el}=0.02, \omega_c=50$ cm$^{-1}$) and strong ($\gamma^{\rm el}=0.7, \omega_c=350$ cm$^{-1}$) electronic dephasing for $\Delta E=200$ cm$^{-1}, \Omega=500$ cm$^{-1}$, and  $T=300$ K. }
\end{center}
\end{figure}

\end{document}